\documentclass[aps,prl,reprint,noaffiliation,groupedaddress,amsmath,amssymb]{revtex4-2}
\usepackage{graphicx}
\usepackage{float}
\usepackage[x11names]{xcolor}
\usepackage[colorlinks,citecolor=red]{hyperref}
\usepackage[all]{hypcap}
\usepackage[T1]{fontenc}

\begin{document}

	\title{Meron Spin Textures in Momentum Space Spawning from Bound States in the Continuum}

        \author{Lixi Rao$^{1}$}
        \author{Jiajun Wang$^{1}$}
        \email{jiajunwang@fudan.edu.cn}
        \author{Xinhao Wang$^{1}$}
        \author{Shunben Wu$^{1}$}
        \author{Xingqi Zhao$^{1}$}
        \author{Wenzhe Liu$^{1,2}$}
        \author{Rensheng Xie$^{5,6}$}
        \author{Yijie Shen$^{5,6}$}
        \email{yijie.shen@ntu.edu.sg}
        \author{Lei Shi$^{1,2,3,4}$}
        \email{lshi@fudan.edu.cn}
        \author{Jian Zi$^{1,2,3,4}$}
        \email{jzi@fudan.edu.cn}

	\affiliation{$^{1}$State Key Laboratory of Surface Physics, Key Laboratory of Micro- and Nano-Photonic Structures (Ministry of Education) and Department of Physics, Fudan University, Shanghai 200433, China}
	\affiliation{$^{2}$Institute for Nanoelectronic Devices and Quantum Computing, Fudan University, Shanghai 200438, China}
	\affiliation{$^{3}$Collaborative Innovation Center of Advanced Microstructures, Nanjing University, Nanjing 210093, China}
	\affiliation{$^{4}$Shanghai Research Center for Quantum Sciences, Shanghai 201315, China}
        \affiliation{$^{5}$Centre for Disruptive Photonic Technologies, School of Physical and Mathematical Sciences, Nanyang Technological University, Singapore 637371, Singapore}
        \affiliation{$^{6}$School of Electrical and Electronic Engineering, Nanyang Technological University, Singapore 639798, Singapore}

	\begin{abstract}
    Topological spin textures, such as merons and skyrmions, have shown significance in both fundamental science and practical applications across diverse physical systems. The optical skyrmionic textures in real space have been extensively explored, but those in momentum space are still rarely studied. Here, we report the experimental generation of momentum-space meron spin textures via bound states in the continuum (BICs) in photonic crystal slabs. We show that under circularly polarized illumination, the momentum-space vortex topology of BICs can transform light with meron spin textures in momentum space. These merons exhibit polarity-switchable configurations controlled by incident light polarization. We theoretically and experimentally verify the generation of momentum-space merons and demonstrate their operational flexibility across a broad spectral range. Our results establish a connection between different momentum-space topologies and provide a robust and compact platform for generating topological spin textures.
	\end{abstract}
	
	\maketitle
    Topological spin textures, describing intricate spin configurations with non-trivial topological numbers, have garnered substantial attention in diverse physical systems. The spin textures can be classified by homotopy groups $\pi_m(S^n)$ describing fields of unit vectors with $n$-degrees of freedom within $m$-dimensional manifolds\cite{manton2004topological}. As 2-dimensional topological spin textures, skyrmionic textures such as skyrmions and merons are initially explored in particle and condensed matter physics, including nucleons, Bose-Einstein condensates, and quantum Hall ferromagnetic states\cite{zhou2025topological}. Carrying robust skyrmion numbers, these skyrmionic spin textures show vital importance in both fundamental science and practical applications\cite{nagaosa2013topological,fert2017magnetic,zhang2020skyrmion}. Recently, skyrmionic spin textures are also found to emerge in various optical systems\cite{tsesses2018optical,du2019deep,dai2020plasmonic,gao2020paraxial,guo2021structured,shen2021supertoroidal,lei2021photonic,zdagkas2022observation,he2022towards,liu2022disorder,shi2023advances,lu2023nanoparticle,wang2024topological,dreher2024spatiotemporal,wang2024observation,shen2024optical,chen2025gouy,leiskyrmionic}, encompassing evanescent electromagnetic fields\cite{tsesses2018optical,du2019deep,lei2021photonic}, energy flux distributions\cite{lu2023nanoparticle,wang2024topological}, and spatiotemporal fields\cite{dai2020plasmonic,guo2021structured,shen2021supertoroidal,zdagkas2022observation,dreher2024spatiotemporal,wang2024observation}. These optical spin textures are proposed and demonstrated with potential applications for optical communication\cite{he2024optical,wang2024topological2,MataCerveraXieLiYuRenShenMaier+2025}, quantum processing\cite{ornelas2024non,ornelas2025topological}, and super-resolution sensing and metrology\cite{yang2023spin,wang2024single}. Most existing generation methods of these optical spin textures primarily rely on real-space interference or superposition of light\cite{tsesses2018optical,du2019deep,davis2020ultrafast,dai2020plasmonic,lei2021photonic,liu2022disorder,zdagkas2022observation,shen2022generation,lu2023nanoparticle,he2024optical,wang2024topological,wang2024topological2,dreher2024spatiotemporal,shen2024optical}. For example, skyrmionic spin textures can be achieved by standing wave interference in surface plasmon systems\cite{tsesses2018optical,du2019deep,dai2020plasmonic,lei2021photonic}, interference of counter-propagating vector beams in $4\pi$-focal systems\cite{wang2024topological}, and structured light engineering by spatial light modulators and metasurfaces\cite{shen2022generation,zdagkas2022observation,liu2022disorder,wang2024topological2,he2024optical}. These real-space methods usually require ingenious design or rigorous alignment. In some other work\cite{guo2020meron}, optical spin textures are theoretically discovered to be distributed in photonic bands of a symmetry-broken honeycomb photonic crystal (PhC) slab. These textures manifest as polarization textures of eigen-radiation near $K$ point, being closely associated with the Berry curvature in momentum space. The observation of momentum-space spin textures reveals novel robust topological properties and hidden connections between spin textures and momentum-space topology. However, till now, momentum-space spin textures are still rarely studied.

    In this Letter, we report the meron spin textures in momentum space generated by bound states in the continuum (BICs) in PhC slabs. BICs are discovered as topological singularities in momentum space\cite{hsu2016bound,huang2023resonant,kang2023applications,wang2024optical}. They are both theoretically and experimentally identified as centers of polarization vortices in momentum space, carrying integer topological charges\cite{zhen2014topological,zhang2018observation,doeleman2018experimental}. With intrinsic momentum-space topological configurations, BICs have been applied in active optical systems to realize topologically structured emissions in lasers\cite{kodigala2017lasing,huang2020ultrafast,sang2022topological} and Bose-Einstein condensation\cite{ardizzone2022polariton,gianfrate2024reconfigurable}. Moreover, considering resonances with incident light from far field, momentum-space topology enabled by BICs are further revealed as ideal platforms to induce spin-orbit interaction modulations in passive optical systems\cite{wang2020generating,wang2022spin}. Representative spin-orbit interaction effects such as spin-to-vortex generation\cite{wang2020generating} and spin Hall effect of light\cite{wang2022spin} have been realized via utilizing momentum-space modulations of BICs.
    \begin{figure}[htbp]
		\includegraphics{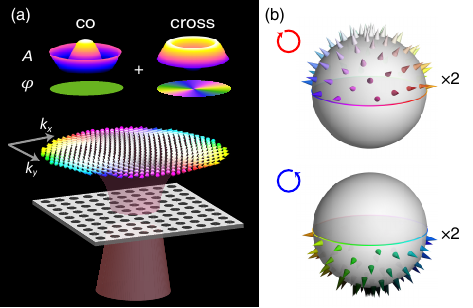}
		\caption{(a) Schematic of meron generations. Merons are generated through resonances with guided modes proximate to BIC at $\Gamma$ point in the PhC slab under circularly polarized illumination. The cross-polarized light carrying vortex phase, combined with co-polarized light, produces polarization fields exhibiting meron spin textures. (b) The generated second-order merons are circular-polarization-dependent and can be mapped onto hemispheres of the Poincar\'e sphere twice. Two types of merons can be switched by changing the circular-polarization state of incident light.}
		\label{Fig1}
    \end{figure}
     These results demonstrate the remarkable capability of BICs for achieving multidimensional light field modulation in momentum space, highlighting the potential of BICs for engineering polarization vortex topologies and exploring other momentum-space topological fields. 
    
    Here, by harnessing momentum-space topology around a BIC, we accomplish the transformation from polarization-vortex topology of BICs to polarity-switchable Stokes meron topology. Through momentum-space imaging techniques\cite{zhang2018observation,wang2020generating}, we experimentally observe two distinct types of second-order meron spin texture, characterized by stable skyrmion numbers across an extensive wavelength range. Our findings extend the horizons of momentum space topology and establish a novel pathway for topological transformations.
    
    Let us start by clarifying the fundamental physical mechanisms underlying the transformation from polarization-vortex topology of BICs to Stokes meron topology. When a monochromatic circularly polarization $|{\sigma_1}\rangle$ shines onto the PhC slab, the transmitted light $|E_{out}\rangle$ can be described as:
    \begin{equation}
			|E_{out}\rangle=t_{1}|\sigma_{1}\rangle+t_{2}|\sigma_{2}\rangle=|t_{1}||{\sigma_1}\rangle+|t_{2}| e^{i\Delta\theta}|{\sigma_2}\rangle.\\
            \label{Eq1}
	\end{equation} 
    Due to the resonance between the incident light and the optical mode of the PhC slab, the transmitted light can have both orthogonal circular-polarization components ($|{\sigma_1}\rangle$ and $|{\sigma_2}\rangle$), corresponding to the co-polarized and cross-polarized components. $t_1$ and $t_2$ are corresponding transmission coefficients. And $\Delta\theta$ characterizes the phase difference between co-polarized and cross-polarized components. For resonances around a BIC, the existed momentum-space polarization vortex will induce the cross-polarized light with a non-zero orbital angular momentum (OAM) charge $l$ in momentum space\cite{wang2020generating}, as illustrated in top panel of Fig. \ref{Fig1}(a). In contrast, the co-polarized light carries a zero OAM charge. Then we can further write the transmitted light as:
    \begin{equation}
			|E_{out}\rangle=A_{1}|{0}\rangle|{\sigma_1}\rangle+A_{2}|{l}\rangle|{\sigma_2}\rangle,\\
            \label{Eq2}
	\end{equation}
    where $A_1$ and $A_2$ refer to the momentum-space amplitude distribution of two circular-polarization components. $|{0}\rangle$ and $|{l}\rangle$ correspond to the OAM charges in the momentum-space phase distribution $\varphi$. From Eq. (\ref{Eq2}), we can find that the total transmitted light $|E_{out}\rangle$ via the momentum-space polarization vortex around the BIC exhibits the representation formulism of the Stokes-type skyrmionic spin textures\cite{gao2020paraxial,he2024optical,MataCerveraXieLiYuRenShenMaier+2025}. In other words, the Stokes-type skyrmionic spin textures in momentum space are expected in the transmitted light. Detailed descriptions of the theoretical model are in the Supplementary Material.
    
    To explore more details, we continue to focus on amplitude and phase distributions of two circular-polarization components of the transmitted light, as shown in top panel of Fig. \ref{Fig1}(a). For the amplitude distribution in momentum space, there is a topologically-protected zero amplitude of $A_2$ at $\Gamma$ point caused by the BIC, and $A_1$ is nonzero due to the direct non-resonant transmission. Considering resonances around the BIC, ratio $A_2/A_1$  increases monotonically from the interior to the exterior within a certain range in momentum space. For the phase distribution in momentum space, the nonzero OAM charge $l$ of the cross-polarized light is directly connected with the topological charge $q$ of the BIC with the relation $|l|=2|q|$\cite{wang2020generating}. Hence, the high-order momentum-space meron spin texture could be confirmed to exist in the transmitted light via the momentum-space polarization vortex around the BIC. For example, the middle panel of Fig. \ref{Fig1}(a) exhibits a schematic of the second-order meron spin texture in momentum space. Notably, the OAM generation via polarization vortex around the BIC is circular-polarization-dependent, i.e. the OAM charge $l$ in cross-polarized light can have opposite sign under different incident circular polarizations. This also indicates momentum-space meron spin textures in transmitted light exhibit circular-polarization-dependent dual configurations and can be modulated by controlling the incident circular polarizations. Explicitly, such Stokes-type meron spin texture can be mapped onto the upper or lower hemispheres of the Poincar\'e sphere with double coverage, corresponding to the incidence of different circular polarizations, respectively, as illustrated in Fig. \ref{Fig1}(b).
    
    \begin{figure}[t]
		\includegraphics{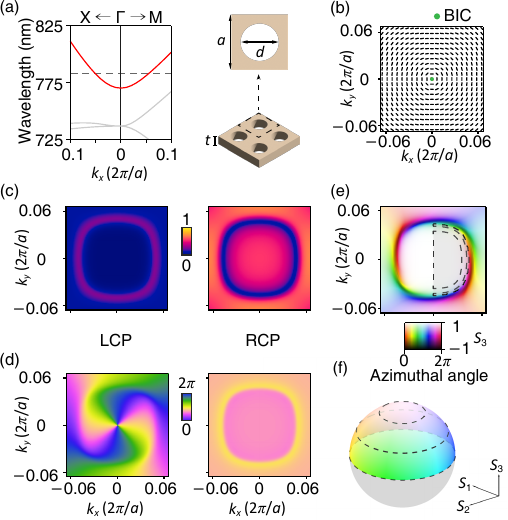}
		\caption{(a) Left: Calculated band structure along $\Gamma$-X and $\Gamma$-M directions. Right: Diagram of designed PhC slab. (b) Momentum-space polarization vortex topology around the BIC at $\Gamma$ point of the highlighted band. (c)-(d) Simulated conversion efficiencies (c) and phase distributions (d) of the transmitted cross-polarized (left) and co-polarized (right) components in momentum space under RCP illumination. (e) Simulated second-order momentum-space meron spin texture of transmitted light. (f) Projected half of the second-order meron on the Poincar\'e sphere. Three dashed lines in (e) and (f) correspond to $S_3$ = 0.9, $S_3$ = 0.5, and $S_3$ = 0.}
		\label{Fig2}
    \end{figure}
    
    To verify the physical pictures above, we designed a PhC slab and studied the generation of momentum-space merons through numerical simulations, as shown in Fig. \ref{Fig2}. The designed PhC slab consists of a freestanding silicon nitride layer (refractive index = 2) patterned with cylindrical holes in a square lattice, as illustrated in right panel of Fig. \ref{Fig2}(a). The structural parameters comprise a thickness $t$ = 100 nm, a hole diameter $d$ = 420 nm, and a lattice period $a$ = 620 nm. The calculated band structure along $\Gamma$-X and $\Gamma$-M directions is shown in left panel of Fig. \ref{Fig2}(a), where the focused band with an at-$\Gamma$ BIC is highlighted in red. We selected an operational wavelength of 782 nm, as indicated by the dashed line in left panel of Fig. \ref{Fig2}(a). The calculated momentum-space polarization field of the highlighted band reveals a BIC with a $+1$ topological charge centered at the $\Gamma$ point, surrounded by a polarization vortex, as shown in Fig. \ref{Fig2}(b).

    Figures \ref{Fig2}(c) and \ref{Fig2}(d) present the momentum-space conversion efficiencies and phase distributions of the transmitted light under the incidence of right-handed circular polarization (RCP). The conversion efficiencies of two circular-polarization components indicate that the intensity ratio increases monotonically from the interior to the exterior within a certain range in momentum space. The cross-polarized component (left-handed circular polarization, LCP) exhibits a phase vortex of $-4\pi$ winding phase, corresponding to $-2$ OAM charge ($|{-2}\rangle$). In contrast, the co-polarized component (RCP) exhibits zero OAM charge ($|{0}\rangle$).  For the total transmitted light with two circular-polarization components, the corresponding spin texture is directly mapped in Fig. \ref{Fig2}(e). Bounded by the normalized third Stokes parameter $S_3$ = 0, a second-order momentum-space meron is observed. As depicted in Fig. \ref{Fig2}(f), half of the meron [gray area in Fig. \ref{Fig2}(e)] is projected comprehensively onto the upper hemisphere of the Poincar\'e sphere. Another type of meron emerges by switching to LCP illumination (see Supplemental Materials).
    
    The boundaries of merons ensure the existence of a well-defined skyrmion number $N_{sk}$, quantified through\cite{guo2020meron,shen2024optical,wang2024topological2}:
    \begin{equation}
    \begin{split}
			N_{sk}&=\frac{1}{4\pi}\iint{\boldsymbol{\mathrm{S}}\cdot\left(\frac{\partial{\boldsymbol{\mathrm{S}}}}{\partial{k_x}}\times\frac{\partial{\boldsymbol{\mathrm{S}}}}{\partial{k_y}}\right)\mathop{}\!\mathrm{d}{k_x}\mathop{}\!\mathrm{d}{k_y}}=\frac{1}{2}{p}\cdot{v}.
            \label{Eq3}
            \end{split}
    \end{equation}	
   Here, $\boldsymbol{\mathrm{S}}$ is the normalized Stokes vector, $p$ is the polarity, and $v$ is the vorticity of the skyrmionic texture. Note that `$\frac{1}{2}$' indicates the ratio of the mapped area of a meron to the Poincar\'e sphere. The polarity is determined by the central Stokes vector direction, which switches between $1$ (up) and $-1$ (down) under RCP or LCP illuminations, respectively. The vorticity characterizes the rotational behavior of the in-plane components of $\boldsymbol{\mathrm{S}}$. The sign of vorticity denotes the rotation direction, positive for counterclockwise and negative for clockwise, and the value of vorticity describes the winding number along a counterclockwise loop around the center\cite{guo2020meron,shen2024optical}, which can be defined as:
    \begin{equation}
    \begin{split}
	      v&=\frac{1}{2\pi}\oint\nabla_{\boldsymbol{k_{||}}}\alpha(\boldsymbol{k_{||}})\cdot\mathrm{d}\boldsymbol{k_{||}}\\
          &=\frac{1}{2\pi}\oint\nabla_{\boldsymbol{k_{||}}}[\varphi_R(\boldsymbol{k_{||}})-\varphi_L(\boldsymbol{k_{||}})]\cdot\mathrm{d}\boldsymbol{k_{||}}=\mp{l},\\
            \label{Eq4}
            \end{split}
    \end{equation}
    where $\alpha(\boldsymbol{k_{||}})$ refers to azimuthal angle of $\boldsymbol{\mathrm{S}}$ on the Poincar\'e sphere, being equal to the phase difference $[\varphi_R(\boldsymbol{k_{||}})-\varphi_L(\boldsymbol{k_{||}})]$ between the RCP and LCP components. And ``$-$" and ``$+$" correspond to RCP and LCP illuminations, respectively. Considering the relationship between OAM charge $l$ and topological charge of BICs $q$ ($l=\mp{2q}$, see Supplemental Materials for detailed derivations), the skyrmion number of generated merons follows: 
    \begin{equation}
	      N_{sk}=\frac{1}{2}{p}\cdot{v}=\pm{q},\\
            \label{Eq5}
    \end{equation}	 
    where ``$+$" and ``$-$" correspond to RCP and LCP illuminations, respectively. Therefore, we can control the generated merons by manipulating the BIC and the incident polarization. More simulated examples are given in the Supplemental Materials.
        
    \begin{figure*}[t]
        \centering
		\includegraphics{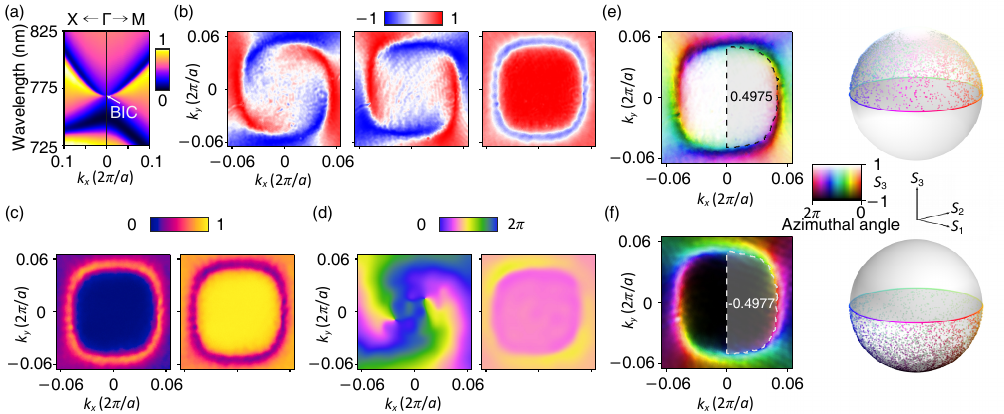}
		\caption{(a) Measured angle-resolved transmittance spectra along $\Gamma$-X and $\Gamma$-M directions. (b) Measured normalized Stokes parameters $S_1$ (middle), $S_2$ (left), and $S_3$ (right) of the transmitted light under RCP illumination. (c)-(d) Experimental intensity proportions (c) and phase distributions (d) of the transmitted cross-polarized (left) and co-polarized (right) components in momentum space under RCP illumination. (e)-(f) Measured spin texture (left) under RCP (e) and LCP (f) illumination. Half of the meron spin textures are mapped onto Poincar\'e spheres (right), where the skyrmion numbers of the half of the merons exhibited within dashed boundaries.}
		\label{Fig3}
    \end{figure*}
    
    To experimentally validate the second-order momentum-space meron spin textures, we fabricated the designed PhC slab and performed measurements using a momentum-space optical measurement system. Figure \ref{Fig3}(a) shows the s-polarized angle-resolved transmittance spectra along $\Gamma$-X and $\Gamma$-M directions, where the white arrow marks the at-$\Gamma$ BIC. By using RCP illumination at 782 nm, the normalized Stokes parameters in momentum space are measured, as shown in Fig. \ref{Fig3}(b). Right panel of Fig. \ref{Fig3}(b) reveals the continuous boundary of the generated momentum-space merons, demarcated by a clear contour where $S_3$ = 0. Measured momentum-space intensity proportions and phase distributions are exhibited in Figs. \ref{Fig3}(c) and \ref{Fig3}(d). Then, the corresponding momentum-space spin texture is illustrated in Fig. \ref{Fig3}(e). For the observed second-order meron spin texture, we calculated a skyrmion number of half of the meron (region encircled by the dashed line) and mapped the projection on the Poincar\'e sphere. For the case of RCP illumination, the skyrmion number is 0.4975 and projection refers to a full coverage on the upper hemisphere of the Poincar\'e sphere. In contrast, Fig. \ref{Fig3}(f) exhibits the case of LCP illumination, in which half the momentum-space meron refers to a skyrmion number of $-0.4977$ and the projection of full coverage on the lower hemisphere of the Poincar\'e sphere. All these results show good accordance with the proposed physical picture and simulations. Here, polarity-switchable momentum-space meron spin textures with circular-polarization-dependent skyrmion numbers generated by BICs have been experimentally demonstrated.
    
    \begin{figure}[b]
		\includegraphics{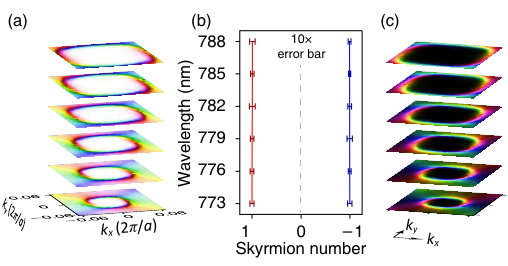}
	    \caption{(a) Second-order merons measured at different wavelengths under RCP illumination. As wavelength increases, second-order merons evolve with the configuration of the band in momentum space. (b) Skyrmion numbers of measured second-order merons (red, RCP; blue, LCP) with $10\times$ magnified error bars. (c) Second-order merons measured at different wavelengths under LCP illumination. The wavelengths of (a) and (c) correspond to (b).}
	\label{Fig4}
    \end{figure}
     
    Figure \ref{Fig4} exhibits the measured momentum-space meron spin textures at a series of wavelengths. Figures \ref{Fig4}(a) and (c) exhibit cases under RCP and LCP illuminations, respectively. As wavelength increases, the meron spin textures evolve with the intersected configuration of the photonic band, exhibiting spatial expansion in momentum space. Figure \ref{Fig4}(b) plots corresponding skyrmion numbers at different wavelengths. And we can see that the skyrmion numbers are stably maintained with values approximately equal to 1 and $-1$ for two illuminations (detailed values of skyrmion numbers provided in Supplemental Materials). The spectral stability of the meron spin textures is due to both the momentum-space topological polarization configurations of BICs and the dispersion of photonic bands. The momentum-space topological polarization configurations of BICs are inherent in the photonic band. During resonances between optical modes in photonic bands and the free-space light field, such topological polarization configurations enable the generation of momentum-space meron spin textures in the transmitted light. Considering the dispersion of photonic bands in PhC slabs, the resonance-based topological generation can also persist across an extensive wavelength range. These results reveal additional wavelength degrees of freedom and also provide considerable operational flexibility for the meron generations.
    
    Topological spin textures in light fields have demonstrated significant potential in various applications\cite{liu2022disorder,yang2023spin,ornelas2025topological,he2024optical,wang2024topological2,wang2024unlocking,ornelas2024non}. For example, quantum skyrmions constructed by two entangled bases exhibit exceptional topological resilience and noise rejection capabilities, showing compelling prospects for quantum communication and quantum computing\cite{ornelas2025topological,ornelas2024non}. In optical communication, the generation of optical skyrmionic textures with multiple configurations enables topology-protected mode division multiplexing strategies\cite{he2024optical,wang2024topological2,MataCerveraXieLiYuRenShenMaier+2025}. In optical computing, high-quality on-chip skyrmion number generation and manipulation provide possibilities for perturbation-resistant integer operations\cite{wang2024unlocking}. In exploring applications of optical skyrmionic textures, compact and efficient generation methods are in high demand. Our BIC-based approach provides a promising platform through compact PhC slabs that require only periodic array fabrications and no real-space alignment, enabling high-quality optical meron generation with multiple configurations and wavelengths.
    
    In summary, we demonstrate meron spin textures in momentum space through a topological transformation mechanism which utilizes the momentum-space vortex topology of BICs in PhC slabs. By switching circular polarizations of incident light, we demonstrate the generation of two distinct meron configurations. This mechanism offers operational flexibility across wavelengths and shows robustness as periodic PhC slabs require no real-space alignment. This robust and compact flat-optics approach to meron generation and manipulation offers promising avenues for photonic applications. Our work provides new insights into on-chip topological spin-optics, bridges different momentum-space topological configurations, and advances the understanding of momentum-space topological transformation. Further research is expected to reveal more hidden aspects of momentum-space topology. Owing to rich topological characteristics, these optical skyrmionic textures also merit further investigation for a range of potential applications, including sensing, metrology, neuromorphic computing, and information processing.

	\bigskip
	
	\begin{acknowledgments}
		This work is supported by National Key R\&D Program of China (2023YFA1406900 and 2022YFA1404800); National Natural Science Foundation of China (No. 12234007, No. 12321161645, No. 12404427, No. 124B2084, No. 12221004, No. T2394480, and No. T2394481); Science and Technology Commission of Shanghai Municipality (24YF2702400, 22142200400, 21DZ1101500, 2019SHZDZX01, 23DZ2260100, and 24142200100). J. W. is also supported by the China National Postdoctoral Program for Innovative Talents (BX20230079) and the China Postdoctoral Science Foundation (2023M740721). Y. S. acknowledges Singapore Ministry of Education (MOE) AcRF Tier 1 grants (RG157/23 \& RT11/23), Singapore Agency for Science, Technology and Research (A*STAR) MTC Individual Research grants (M24N7c0080), and Nanyang Assistant Professorship Start Up grant.
		
		\bigskip
		
		L. R. and J. W. contributed equally to this work.
		
	\end{acknowledgments}

\end{document}